\documentclass{article}

\usepackage{arxiv}
\usepackage{times}

\usepackage{soul}
\usepackage{url}
\usepackage[utf8]{inputenc}
\usepackage[small]{caption}
\usepackage{graphicx}
\usepackage{amsmath}
\usepackage{booktabs}
\usepackage{latexsym} 
\usepackage{amssymb}
\usepackage{amsthm}
\usepackage[group-separator={,}]{siunitx}
\usepackage{algorithm}
\usepackage[noend]{algpseudocode}
\usepackage{complexity}
\usepackage{multirow}
\usepackage{caption}
\usepackage{verbatim}
\usepackage{enumitem}
\usepackage{subfigure}
\usepackage{balance}
\usepackage{dblfloatfix}
\usepackage{url}

\newtheorem{mytheorem}{Theorem}

\newtheorem{mycorollary}{Corollary}
\newtheorem{mydefinition}{Definition}
\newtheorem{myexample}{Example}

\newtheorem{myobservation}{Observation}

\newcommand{\myqed}{\mbox{$\diamond$}}

\newcolumntype{C}{>{\centering\arraybackslash}p{4cm}}
\DeclareMathOperator*{\argmax}{arg\,max}
\DeclareMathOperator*{\argmin}{arg\,min}

\title{Envy-freeness up to one item: \\ Shall we add or remove resources?}

\author{
Martin Aleksandrov
\affiliations
TU Berlin, Germany
\emails
martin.aleksandrov@tu-berin.de
}

\begin{document}

\maketitle

\begin{abstract}
We consider a fair division model in which agents have general valuations for bundles of indivisible items. We propose two new axiomatic properties for allocations in this model: EF1$^{\pm}$ and EFX$^{\pm}$. We compare these with the existing EF1 and EFX. Although EF1 and EF1$^{\pm}$ allocations often exist, our results assert eloquently that EFX$^{\pm}$ and PO allocations exist in each case where EFX and PO allocations do not exist. Additionally, we prove several new impossibility and incompatibility results.
\end{abstract}

\section{Introduction}\label{sec:intro}

We study fair division problems where agents have \emph{general} (i.e.\ positive, zero or negative) valuations for bundles of indivisible items. Some items are marginally advantageous for agents. We call these goods. Item $o$ is a \emph{good} for agent $i$ with respect to bundle $M$ if the $i$'s marginal valuation for $o$ when added to $M$ is non-negative. Other items are marginally disadvantageous for agents. We call these bads. Item $o$ is a \emph{bad} for agent $i$ with respect to bundle $M$ if the $i$'s marginal valuation for $o$ when added to $M$ is non-positive. We consider three item types depending on combinations of valuations. 

We refer to item $o$ as \emph{mixed} if, there is a pair of agents $i,j$ and a pair of disjoint bundles $M,N$ such that the marginal valuation of $i$ for $o$ when added to $M$ is strictly positive and the marginal valuation of $j$ for $o$ when added to $N$ is strictly negative. Also, we refer to item $o$ as \emph{generally good} for agent $i$ if $o$ is good for $i$ wrt any bundle $M$. Likewise, we refer to item $o$ as \emph{generally bad} for agent $i$ if $o$ is bad for $i$ wrt any bundle $M$. Thus, we consider three problem types depending on available items.

The first type of problems has only items that are \emph{generally good} and {\em generally bad} for agents. In other words, each agent consider a given item as generally good or generally bad. From a theoretical perspective, this could be the case whenever we combine a problem with goods and a problem with bads. From a practical perspective, any problem where the valuations are additive (i.e.\ the agent's valuation for a bundle of items is the sum of their valuations for the items in the bundle) has only generally good items and generally bad items. Popular such applications are paper assignments, food allocations, course allocations, etc. 

Although such problems capture various common settings, there are other practical settings where none of the items is generally good/bad for anyone. We show this in Example~\ref{exp:gen}.

\begin{myexample}\label{exp:gen}
Alice and Bob love playing cricket but they have only one ball $b$ and one racket $r$. Let them value $\lbrace b,r\rbrace$ at $2$, $\lbrace b\rbrace$ and $\lbrace r\rbrace$ at $-1$, and $\emptyset$ at $0$. Giving Alice $b$ in addition to $r$ makes her happy because she gets utility $2$ but giving her only $b$ makes her unhappy because she gets utility $-1$. This means that her judgment about whether $b$ is good or bad depends on whether she receives $r$ or not. Hence, neither $b$ nor $r$ is considered as generally good/bad by Alice or Bob. 
\myqed
\end{myexample}

This leads us to the other two problem types. The second type has {\em no} mixed items and capture settings where all agents reach a consensus about whether a given item is good or bad with respect to their bundles. For instance, a group of friends all agree to go on a holiday or not given their budgets. The third type has items that could be \emph{arbitrary} (i.e.\ possibly mixed) and capture settings such as the one in Example~\ref{exp:gen}. Both types model complementarities or substitutabilities. Say, I value the right and left shoes together and have no value for just one shoe. Also, I value my bicycle but disvalue a second one due to the lack storage space in my basement. 

\begin{figure}[h]
\centering
\includegraphics[scale=0.235]{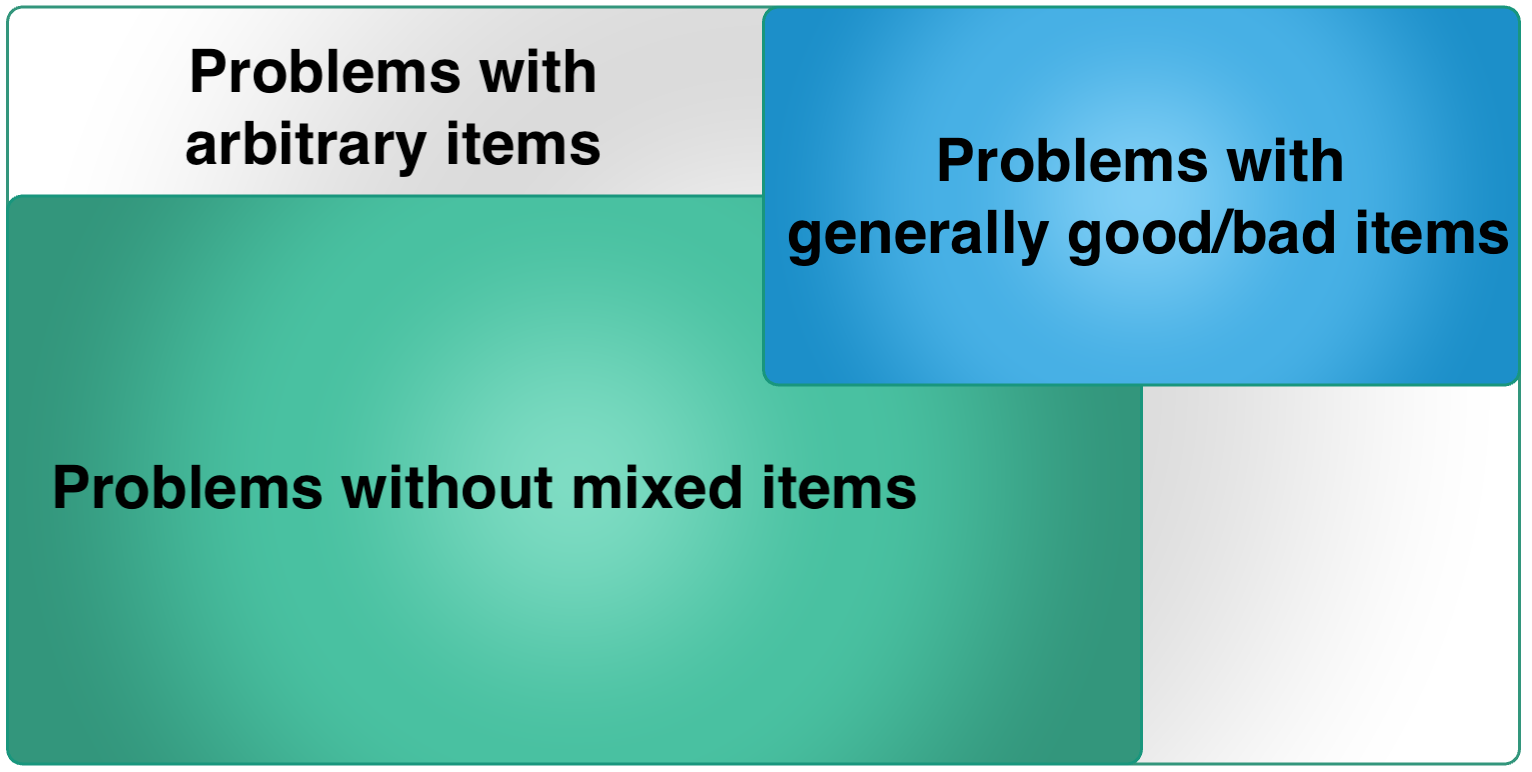}
\caption{Taxonomy of problems with general valuations.}
\label{fig:taxgen}
\end{figure}

\begin{figure}[h]
\centering
\includegraphics[scale=0.235]{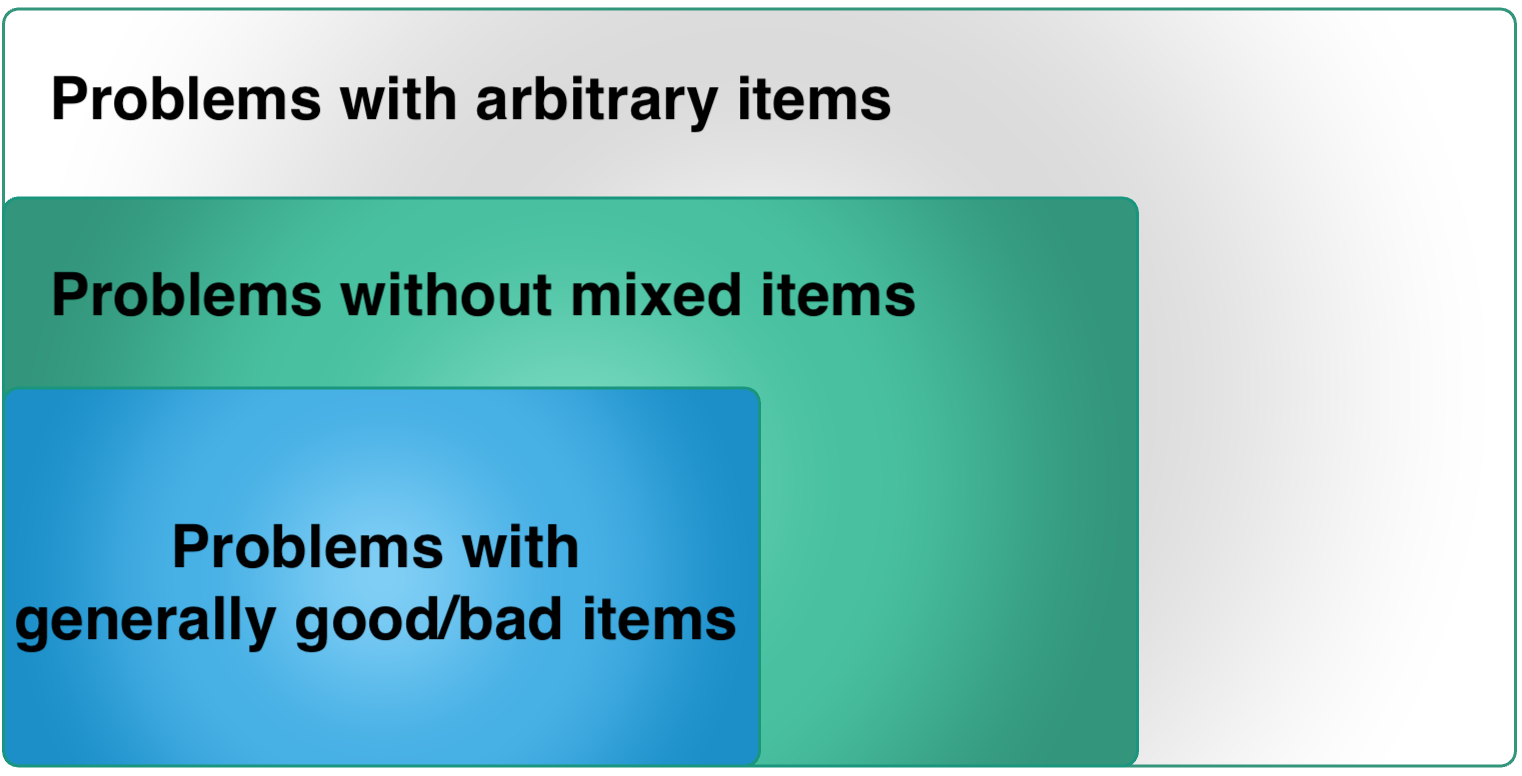}
\caption{Taxonomy of problems with identical general valuations.}
\label{fig:taxident}
\end{figure}

An interesting case in our setting is when the agents' valuations are general but \emph{identical} (i.e.\ for each bundle, the agents' valuations are equal). For example, people tend to have a similar value for a pair of Nike shoes but perhaps this value differs from their value for a pair of Adidas shoes. Further, computer science students at TU Berlin tend to value one module (i.e.\ a set of courses) identically (e.g.\ by means of ECTS credits) but they may value differently another module. We will shortly observe that the three sets of problems relate as in Figures~\ref{fig:taxgen} and~\ref{fig:taxident}.

An (complete) allocation exhausts all items by giving to each agent a different bundle. For allocations in these problems, we study approximations of the golden standard in fair division: \emph{envy-freeness} (i.e.\ no agent envies another one) \cite{foley1967}. Two such existing properties for our setting are EF1 and EFX \cite{aziz2019gc}. For example, EFX requires that, if agent $i$ envies agent $j$, 1) \emph{removing} any item from agent $i$'s bundle, that is non-zero-valued bad for agent $i$ wrt agent $i$'s bundle, makes agent $i$ envy-free of agent $j$, and 2) \emph{removing} any item from agent $j$'s bundle, that is non-zero-valued good for agent $i$ wrt agent $j$'s bundle, makes agent $i$ envy-free of agent $j$. EF1 is a weaker property that relaxes the EFX conditions to some (possibly zero-valued) and not any (non-zero-valued) item.

We propose in our work two alternative properties. Both of them aim at restoring envy-freeness by reducing the valuations of envied agents. This can happen either through removing goods from their bundles or adding bads to these bundles. For this purpose, we refer to them as EFX$^{\pm}$ and EF1$^{\pm}$. For example, EFX$^{\pm}$ requires that, if agent $i$ envies agent $j$, 1)$^\prime$ \emph{adding} to agent $j$'s bundle any item from agent $i$'s bundle, that is non-zero-valued bad for agent $i$ wrt agent $i$'s bundle, makes agent $i$ envy-free of agent $j$, and 2)$^\prime$ \emph{removing} any item from agent $j$'s bundle, that is non-zero-valued good for agent $i$ wrt agent $i$'s bundle, makes agent $i$ envy-free of agent $j$. In a similar fashion, we propose to consider a weaker version - EF1$^{\pm}$ - that relaxes the EFX$^{\pm}$ conditions to some item and not any item.

The idea of reducing the valuation of an envied agent by adding to their bundle a bad from an envy agent's bundle is already used in the literature \cite{chen2020fairness}. Further, EFX requires that each removed item from the envied agent's bundle is non-zero-valued good for the envy agent wrt the envied agent' bundle. However, such an item could still be bad for the envy agent wrt their own bundle. For this reason, we believe that this requirement is counter-intuitive as we would expect the envy agent to be happy that such a bad is not in their own bundle but in the envied agent's bundle. In contrast, the {\em novelty} about EFX$^{\pm}$ is that it requires that each removed item from the envied agent's bundle is non-zero-valued good for the envy agent wrt their own bundle.

We also study combinations of each of these fairness properties with efficiency in allocations. A classical efficiency notion is Pareto-optimality (PO) \cite{pareto1896}. Pareto-optimality ensures that we cannot re-distribute items among agents' bundles in such a way so that we make every agent weakly happier and some agent strictly happier. Thus, we compare in our work various combinations of properties.

For example, we will observe several similarities between the new EF1$^{\pm}$, EFX$^{\pm}$ and the existing EF1, EFX. One such is for the additive case. In this case, an allocation is EFX (EF1) iff it is EFX$^{\pm}$ (EF1$^{\pm}$). Furthermore, EF1 and EF1$^{\pm}$ allocations exist in problems with \num{2} agents and general (not just identical) valuations. We will also prove some major differences between the new and the existing fairness properties. We remark that EFX might be unachievable and it is unknown whether EF1 is achievable in problems with identical general valuations \cite{brczi2020envyfree}. In contrast, an EFX $^{\pm}$ (and, hence, EF1$^{\pm}$) and PO allocation exists in this and other cases. We give further motivation via an example.

\section{Motivation}\label{sec:mot}

Many applications of fair division in practice are concerned with issues such as privacy and anonymity of the participants taking part in the allocation. Typical examples are school choice, conference paper assignment, voting, task assignment and course allocation problems. 

Indeed, we may want a system that does not reveal knowledge about the preferences to the public. At TU Berlin, it has been observed that this increases the dissatisfaction of both students and staff members because it generates jealousy and complaints. We next capture this in Example~\ref{exp:mot}.

\begin{myexample}\label{exp:mot}
Consider students X and Y who wish to enrol in university courses $1$, $2$, $3$ and $4$, supposing there is one empty spot in each course. Let $1$ be a seminar class and $2,3,4$ lecture classes. We let the students value a module of $1$ course with $6$ ECTS credits, a module of $3$ courses with $12$ ECTS credits and a module of $4$ courses with $18$ ECTS credits. 

Furthermore, we let the valuations of students for modules of $2$ courses depend on whether these are lectures and/or seminars. Suppose that combining the seminar class with any lecture class into a module does not give additional credits to the students (i.e.\ $6$ credits) while combining two lecture classes into a module gives $3$ additional credits (i.e.\ $9$ credits).

Assigning X to the seminar class and Y to the three lecture classes gives $6$ credits to X and $12$ credits to Y. This outcome is EFX$^{\pm}$ and PO, but not EFX. By comparison, assigning X to the seminar class and a lecture class and Y to two lecture classes gives $6$ credits to X and $9$ credits to Y. This outcome is EFX and EFX$^{\pm}$ but not PO.
\myqed
\end{myexample}

If the central planner reveals the preferences of X and Y, then X would be able to compare their valuations with those of student Y. For this reason, the EFX allocation in which each of X and Y is assigned to a module of \num{2} courses might be preferred. But, this outcome is not PO.

Otherwise, X would not be able to compare their valuations with the valuations of Y. As a result, the EFX$^{\pm}$ allocation in which X is assigned to a module of \num{1} course and Y is assigned to a module of \num{3} courses might be preferred. Additionally, this one satisfies PO. 

Notably, achieving EFX$^{\pm}$ guarantees that each student get at least as much as they would get when achieving EFX. This is because EFX$^{\pm}$ can often be combined with PO. Indeed, the valuations of each of X and Y in the EFX$^{\pm}$ allocation are at least as much as their valuations in the EFX allocation.

\begin{table*}[t]
\resizebox{\textwidth}{!}{
\centering
\begin{tabular}{|c|c|C|C|C|C|C|C|}
\hline

\multirow{2}{*}{property} & \multirow{2}{*}{agents} & \multicolumn{2}{C|}{problems with} & \multicolumn{2}{C|}{problems without} & \multicolumn{2}{C|}{problems with} \\ 

& & \multicolumn{2}{C|}{arbitrary items} & \multicolumn{2}{C|}{mixed items} & \multicolumn{2}{C|}{gen. good/bad items} \\ \hline

& & \multicolumn{2}{c}{} & \multicolumn{2}{c}{\bf identical general valuations} & \multicolumn{2}{c|}{} \\ \hline

EFX & $\geq 2$ &\multicolumn{2}{c|}{$\times$, (Thm~\ref{thm:impefx})} & \multicolumn{2}{c|}{open} & \multicolumn{2}{c|}{open} \\ \hline

EFX ($\exists$) \& EFX$^{\pm}$ & $\geq 2$ &\multicolumn{2}{c}{} & \multicolumn{2}{c}{} & \multicolumn{2}{c|}{$\times$, (Thm~\ref{thm:noefxstronger})} \\ \hline

EFX ($\exists$) \& PO & $\geq 2$ & \multicolumn{2}{c}{} & \multicolumn{2}{c}{} & \multicolumn{2}{c|}{$\times$, (Thm~\ref{thm:noefxpostronger})} \\ \hline

EFX$^{\pm}$ \& PO & $\geq 2$ &\multicolumn{2}{c}{$\checkmark$, leximin (Thm~\ref{thm:newefxpo})} & \multicolumn{2}{c}{} & \multicolumn{2}{c|}{} \\ \hline

& & \multicolumn{2}{c}{} & \multicolumn{2}{c}{\bf general valuations} & \multicolumn{2}{c|}{} \\ \hline

EFX$^{\pm}$ & $2$ & \multicolumn{2}{c}{$\checkmark$, (Cor~\ref{cor:one})} &  \multicolumn{2}{c}{} &  \multicolumn{2}{c|}{} \\ \hline

EFX$^{\pm}$ \& PO & $2$ & \multicolumn{2}{c}{} &  \multicolumn{2}{c}{} &  \multicolumn{2}{c|}{$\times$ (Cor~\ref{cor:two})} \\ \hline

EFX$^{\pm}$ \& PO (d.n.v.) & $2$ & \multicolumn{2}{c}{$\checkmark$, leximin (Thm~\ref{thm:newefxpotwo})} &  \multicolumn{2}{c}{} &  \multicolumn{2}{c|}{} \\ \hline

EF1 & $2$ &  \multicolumn{2}{c}{$\checkmark$, \cite{brczi2020envyfree}} &  \multicolumn{2}{c}{} &  \multicolumn{2}{c|}{} \\ \hline

EF1 \& PO & $2$ &  \multicolumn{2}{c|}{$\times$, (Thm~\ref{thm:noefonepo})} &  \multicolumn{2}{c|}{open} &  \multicolumn{2}{c|}{open} \\ \hline

EF1$^{\pm}$ \& PO & $2$ &  \multicolumn{2}{c|}{$\times$, (Cor~\ref{cor:three})} &  \multicolumn{2}{c|}{open} &  \multicolumn{2}{c|}{open} \\ \hline

EF1$^{\pm}$, EF1 & $\geq 3$ &  \multicolumn{2}{c|}{open} &  \multicolumn{2}{c|}{open} &  \multicolumn{2}{c|}{open} \\ \hline

\end{tabular}
}
\caption{Key: $\checkmark$-possible, $\times$-not possible (non-zero marginal valuations), $\exists$-EFX allocations, d.n.v.-disjointly normalised valuations.}
\label{tab:results}
\end{table*}

\section{Our contributions}\label{sec:con}

We emphasise in our work on possibility and impossibility results. Table~\ref{tab:results} contains our and some existing results. Even though we close interesting major cases, we leave some open questions for future work.

\begin{itemize}
\item For general but identical valuations, an EFX allocation might be unachievable or incompatible (Theorems~\ref{thm:impefx}-~\ref{thm:noefxpostronger}). 
\item For general but identical valuations, an EFX$^{\pm}$ and PO allocation exists (Theorem~\ref{thm:newefxpo}).
\item For general but disjointly normalised valuations and \num{2} agents, an EFX$^{\pm}$ and PO allocation exists (Theorem~\ref{thm:newefxpotwo}).
\item For general valuations, an EF1 and PO allocation might not exist (Theorem~\ref{thm:noefonepo}). We note that this compatibility remains an open problem with additive valuations.
\item Finally, we make additionally several minor contributions (Corollaries~\ref{cor:one}-~\ref{cor:three}). 
\end{itemize} 

We feel that our theoretical results provide further motivation for using EF1$^{\pm}$ and EFX$^{\pm}$ instead of EF1 and EFX. EF1$^{\pm}$ is often achievable when EF1 is, and EFX$^{\pm}$ and PO are achievable in cases when EFX and PO are not. 

\section{Related work}\label{sec:rel}

For indivisible goods, EF1 was proposed by Budish \shortcite{budish2011} and EFX by Caragiannis et al. \shortcite{caragiannis2016}. For our setting, they were generalized by Aziz et al. \shortcite{aziz2019gc}. They gave the double round-robin algorithm for computing EF1 allocations in problems with additive valuations. Plaut and Roughgarden \shortcite{plaut2018} proved that a stronger version of EFX (labelled as EFX$_0$ in \cite{kyropoulou2019}) can be satisfied in problems with general but identical valuations for goods (i.e.\ generally good items). By comparison, we prove that EFX$^{\pm}$ and PO allocations exist in each problem with such valuations for mixed manna. We also give problems with identical general valuations for mixed manna where EFX (and, therefore, EFX and PO) allocations might not exist even when the marginal valuations are non-zero. 

Chen and Liu \shortcite{chen2020fairness} considered problems with generally good/bad items. They proposed the following variant of EFX: if agent $i$ envies agent $j$, {\em adding} to agent $j$'s bundle any item from agent $i$'s bundle, that is {\em generally} bad for agent $i$, makes agent $i$ envy-free of agent $j$, and {\em removing} any item from agent $j$'s bundle, that is {\em generally} good for agent $i$, makes agent $i$ envy-free of agent $j$. In our setting, there are problems where each mixed item is not generally good/bad for anyone (see Example~\ref{exp:gen}). For this reason, this variant is well-defined only for problems with generally good/bad items. They showed that their variant and PO are compatible under the identical and non-zero marginal assumptions. We illustrate that this breaks whenever we drop the latter assumption. Moreover, EFX$^{\pm}$ is well-defined for all problems. 

Very recently, Bérczi et al.\ \shortcite{brczi2020envyfree} showed that EFX (labelled as EFX$^+_-$) allocations may not exist in problems with identical general valuations whose marginals could be zero. From this perspective, our impossibility result about EFX under the non-zero marginal assumption is stronger. They also showed that a stronger version of EFX (i.e.\ EFX$^+_0$) can be satisfied in problems with general but identical valuations for bads (i.e.\ generally bad items). They observed a similar result for the case of \num{2} but dropping the identical assumption. However, we discuss later that adding goods to the problem may destroy a natural generalization of EFX$^+_0$ even when the number of agents is \num{2} and the valuations are identical. Also, none of their results includes PO as some of our results.

Recently, Aleksandrov \shortcite{aleksandrov2020jfx} confirmed that EFX and PO allocations exist in problems with \num{2} agents and normalised (i.e.\ the total valuation is equal) additive valuations for mixed manna. This is an important case because it is practical. For example, some web-applications on Spliddit ask agents to share a pre-defined total valuation for items \cite{caragiannis2016}. We give a similar result for EFX$^{\pm}$ and PO allocations under the assumption of general but disjointly normalised (i.e.\ the valuation of each partition is equal) valuations for mixed manna. We also note that EFX allocations exist in problems with \num{3} agents and additive valuations for goods \cite{chaudhury2020efxthree}. However, the case of \num{4} or more agents in such problems remains open. 

\section{Formal preliminaries}\label{sec:pre}

In this section, we give the formal preliminaries for our analysis: model, problems, properties and solutions. We also confirm the set taxonomies in Figures~\ref{fig:taxgen} and~\ref{fig:taxident}.

\subsection{Model}\label{subsec:model}

We let $[n]$ denote a set of agents and $[m]$ denote a set of indivisible items, where $n,m\in\mathbb{N}_{\geq 2}$. Further, we let each $i\in [n]$ use some function $v_i$ to specify their {\em general} valuation $v_i(M)\in\mathbb{R}$ for each $M\subseteq [m]$. We write $v_i(o)$ for $v_i(\lbrace o\rbrace)$. The valuation $v_i(M)$ is {\em additive} if $v_i(M)=\sum_{o\in M} v_i(o)$. The marginal valuations are \emph{non-zero} if, for each $i\in [n]$, each $o\in [m]$ and each $M\subseteq [m]\setminus\lbrace o\rbrace$, $v_i(M\cup\lbrace o\rbrace)-v_i(M)\neq 0$. The valuations \emph{identical} if, for each $M\subseteq [m]$, $v_i(M)=v_j(M)$ for each $i,j\in [n]$. We write then $v(M)$.

We refer to item $o$ as \emph{mixed} if, there is a pair of agents $i,j$ and a pair $M\subseteq [m]\setminus\lbrace o\rbrace,N\subseteq [m]\setminus(M\cup\lbrace o\rbrace)$ such that $v_i(M\cup\lbrace o\rbrace)>v_i(M)$ and $v_j(N\cup\lbrace o\rbrace)<v_j(N)$ hold. We refer to item $o$ as \emph{good} for agent $i$ wrt $M\subseteq [m]$ if $v_i(M\cup\lbrace o\rbrace)\geq v_i(M)$. We refer to item $o$ as \emph{generally good} for agent $i$ if $o$ is good for $i$ wrt each $M\subseteq [m]$. We refer to item $o$ as \emph{bad} for agent $i$ wrt $M\subseteq [m]$ if $v_i(M\cup\lbrace o\rbrace)\leq v_i(M)$. We refer to item $o$ as \emph{generally bad} for agent $i$ if $o$ is bad for $i$ wrt each $M\subseteq [m]$. 

In a problem \emph{with arbitrary items}, some items could be mixed. In a problem {\em without mixed items}, there are no such items. An item now can be good for everyone in one allocation (i.e.\ their marginal valuations are non-negative) and bad for everyone in another allocation (i.e.\ their marginal valuations are non-positive). Also, some item could be good for everyone and another item could be bad for everyone in the same allocation. In a problem \emph{with generally good/bad items}, an item that is good/bad for some agent in a given allocation is also good/bad for them in any other allocation.

\subsection{Problem taxonomy}\label{subsec:comp}

Example~\ref{exp:gen} is a witness where each item is mixed and no item is generally good/bad for anyone. However, in problems with general valuations, a mixed item can be generally good for one agent and generally bad for another agent (see Figure~\ref{fig:taxgen}).

\begin{myobservation}\label{obs:one}
Even with additive (not necessarily identical) valuations, there are problems with generally good and generally bad items, in which some items are mixed.
\end{myobservation}

\begin{myproof}
Let us consider a problem with \num{2} agents and the valuations: $v_1(a)=3$, $v_1(b)=-1$ and $v_2(a)=1$, $v_2(b)=1$. We focus on item $b$. This item is mixed because of $v_1(\emptyset\cup\lbrace b\rbrace)<v_1(\emptyset)$ and $v_2(\lbrace a\rbrace\cup\lbrace b\rbrace)>v_2(\lbrace b\rbrace)$. However, item $b$ is generally bad for agent 1 - $v_1(b)<v_1(\emptyset)$ and $v_1(\lbrace a,b\rbrace)<v_1(a)$ - and generally good for agent 2 - $v_2(b)>v_2(\emptyset)$ and $v_2(\lbrace a,b\rbrace)>v_2(b)$.
\myqed
\end{myproof}

By comparison, in problems with identical general valuations, an item that is generally good/bad for some agent is also generally good/bad for each other agent. Hence, such an item cannot be mixed (see Figure~\ref{fig:taxident}).

\begin{myobservation}\label{obs:two}
With general but identical valuations, a problem with generally good and generally bad items is also a problem without mixed items.
\end{myobservation}

\begin{myproof}
Let us consider an allocation $A=(A_1,\ldots,A_n)$. Pick some item with non-zero marginal valuation for some agent. Wlog, $o\in A_1$. If $v(A_1)-v(A_1\setminus\lbrace o\rbrace)>0$, then $o$ is good for every agent simply because the valuations are identical. Since the problem is with generally good items, it follows that $o\in G_i$ holds for each $i\in [n]$. But, then $v(A_j\cup\lbrace o\rbrace)-v(A_j)\geq 0$ holds for each $j\in [n]\setminus\lbrace 1\rbrace$. If $v(A_1)-v(A_1\setminus\lbrace o\rbrace)<0$, then $o$ is bad for every agent. Since the problem is with generally bad items, it follows that $o\in B_i$ holds for each $i\in [n]$. But, then $v(A_j\cup\lbrace o\rbrace)-v(A_j)\leq 0$ holds for each $j\in [n]\setminus\lbrace 1\rbrace$. Finally, each remaining item is such that the marginal valuation of any agent for it is zero.
\myqed
\end{myproof}

In a problem with identical additive valuations, there are no mixed items. Also, an item is good/bad for an agent in any allocation. Hence, this is a problem with generally good/bad items. This might be untrue in general (see Figure~\ref{fig:taxident}).

\begin{myobservation}\label{obs:three}
With identical general valuations, there are problems without mixed items, in which some items are not generally good or generally bad for any agent.
\end{myobservation}

\begin{myproof}
Consider \num{2} agents and the valuations $v(\emptyset)=0$, $v(a)=1$, $v(b)=1$, $v(c)=3$, $v(d)=1$, $v(\lbrace a,b\rbrace)=2$, $v(\lbrace a,c\rbrace)=2$, $v(\lbrace a,d\rbrace)=2$, $v(\lbrace b,c\rbrace)=2$, $v(\lbrace b,d\rbrace)=2$, $v(\lbrace c,d\rbrace)=2$, $v(\lbrace a,b,c\rbrace)=4$, $v(\lbrace b,c,d\rbrace)=4$, $v(\lbrace a,b,d\rbrace)=1.5$, $v(\lbrace a,c,d\rbrace)=4$ and $v(\lbrace a,b,c,d\rbrace)=5$.

We note that there are no mixed items in this problem: for each $o\in\lbrace a,b,c,d\rbrace$ and $S,T\subseteq\lbrace a,b,c,d\rbrace\setminus\lbrace o\rbrace$ such that $S\cap T=\emptyset$ and $S\cup T=\lbrace a,b,c,d\rbrace\setminus\lbrace o\rbrace$, at most one of the two holds: (1) $v(S\cup\lbrace o\rbrace)> v(S)$ and $v(T\cup\lbrace o\rbrace)> v(T)$ or (2) $v(S\cup\lbrace o\rbrace)< v(S)$ and $v(T\cup\lbrace o\rbrace)< v(T)$. 

We also note that item $a$ is good in $A=(\lbrace a,b\rbrace,\lbrace c,d\rbrace)$ and bad in $B=(\lbrace a,c\rbrace,\lbrace b,d\rbrace)$: $u_1(A_1)-u_1(A_1\setminus\lbrace a\rbrace)=1>0$, $u_2(A_2\cup\lbrace a\rbrace)-u_2(A_2)=2>0$, $u_1(B_1)-u_1(B_1\setminus\lbrace a\rbrace)=-1<0$ and $u_2(B_2\cup\lbrace a\rbrace)-u_2(B_2)=-0.5<0$. Hence, item $a$ is not generally good/bad for any of the agents.
\myqed
\end{myproof}

\subsection{Axiomatic properties}\label{subsec:axioms}

An \emph{(complete) allocation} $A=(A_1,\ldots,A_n)$ is such that (1) $A_a$ is the bundle of agent $a\in [n]$, (2) $\cup_{a\in [n]} A_a=[m]$ and (3) $A_a\cap A_b=\emptyset$ for each $a,b\in[n]$ with $a\neq b$. We write $\overrightarrow{v}(A)\in\mathbb{R}^n$ for the non-decreasing valuation vector wrt $A$. 

\paragraph{Envy-freeness up to one removed good/removed bad}\label{par:efxrem} We first define the existing approximations EF1 and EFX. 

\begin{mydefinition}$(${\em EF1}$)$
An allocation $A$ is \emph{envy-free up to some removed good or some removed bad} if, $\forall i, j\in [n]$ s.t. $i$ envies $j$, 1) $\exists o\in A_j$ s.t. $v_i(A_i)\geq v_i(A_j\setminus\lbrace o\rbrace)$ or 2) $\exists o\in A_i$ s.t. $v_i(A_i\setminus\lbrace o\rbrace)\geq v_i(A_j)$.
\end{mydefinition} 

\begin{mydefinition}$(${\em EFX}$)$
An allocation $A$ is \emph{envy-free up to any non-zero removed good and any non-zero removed bad} if, $\forall i, j\in [n]$ s.t. $i$ envies $j$, 1) $\forall o\in A_j$ s.t. $v_i(A_j)> v_i(A_j\setminus\lbrace o\rbrace)$: $v_i(A_i)\geq v_i(A_j\setminus\lbrace o\rbrace)$ and 2) $\forall o\in A_i$ s.t. $v_i(A_i)< v_i(A_i\setminus\lbrace o\rbrace)$: $v_i(A_i\setminus\lbrace o\rbrace)\geq v_i(A_j)$.
\end{mydefinition} 

An allocation that satisfies EFX also satisfies EF1. It is well-known that the opposite relation may not be true even with additive valuations.

\begin{center}
EFX\hspace{0.25cm}$\Rightarrow$\hspace{0.25cm}EF1
\end{center}   

\paragraph{Envy-freeness up to one removed good/added bad}\label{par:efonezero} We further define the novel approximations EF1$^{\pm}$ and EFX$^{\pm}$.

\begin{mydefinition}$(${\em EF1$^{\pm}$}$)$
An allocation $A$ is \emph{envy-free up to some removed good or some added bad} if, $\forall i, j\in [n]$ s.t. $i$ envies $j$, 1)$^{\prime}$ if $\exists o\in A_j$ s.t. $v_i(A_i\cup\lbrace o\rbrace)>v_i(A_i)$, then $v_i(A_i)\geq v_i(A_j\setminus\lbrace o\rbrace)$, or 2)$^{\prime}$ if $\exists o\in A_i$ s.t. $v_i(A_i)<v_i(A_i\setminus\lbrace o\rbrace)$, then $v_i(A_i)\geq v_i(A_j\cup\lbrace o\rbrace)$.
\end{mydefinition} 

\begin{mydefinition}$(${\em EFX$^{\pm}$}$)$
An allocation $A$ is \emph{envy-free up to any non-zero removed good and any non-zero added bad} if, $\forall i, j\in [n]$ s.t. $i$ envies $j$, 1)$^{\prime}$ $\forall o\in A_j$ s.t. $v_i(A_i\cup\lbrace o\rbrace)>v_i(A_i)$: $v_i(A_i)\geq v_i(A_j\setminus\lbrace o\rbrace)$ and 2)$^{\prime}$ $\forall o\in A_i$ s.t. $v_i(A_i)<v_i(A_i\setminus\lbrace o\rbrace)$: $v_i(A_i)\geq v_i(A_j\cup\lbrace o\rbrace)$.
\end{mydefinition} 

An allocation that is EFX$^{\pm}$ also satisfies EF1$^{\pm}$. With additive valuations, an allocation is EFX$^{\pm}$ iff it is EFX. Also, it is EF1$^{\pm}$ iff it is EF1. This follows because at least one item in the envy agent's bundle or the unenvied agent's bundle generates an increase in the valuation of the envy agent. Therefore, an allocation that is EF1$^{\pm}$ may violate EFX$^{\pm}$.

\begin{center}
EFX$^{\pm}$\hspace{0.25cm}$\Rightarrow$\hspace{0.25cm}EF1$^{\pm}$
\end{center}   

\paragraph{Pareto-optimality} As we mentioned earlier, we also consider a classical efficiency criterion such as Pareto-optimality. 

\begin{mydefinition}$(${\em PO}$)$
An allocation $A$ is \emph{Pareto-optimal} if there is no allocation $B$ that \emph{Pareto-improves} $A$, i.e.\ $\forall i\in [n]$: $v_i(B_i)\geq v_i(A_i)$ and $\exists j\in [n]$: $v_j(B_j)> v_j(A_j)$. 
\end{mydefinition} 

\subsection{The leximin solution}\label{subsec:leximin}

We consider the {\em leximin} solution from \cite{dubins1961}. This one maximizes the minimum utility of any agent in an allocation, subject to which the second minimum utility is maximized, and so on. 

Plaut and Roughgarden \shortcite{plaut2018} implemented one total operator for comparing allocations: $\succ$. We write $A\succ B$ (i.e.\ $A$ {\em leximin-dominates} $B$) if there exits an index $i\leq n$ such that $\overrightarrow{v}(A)_j=\overrightarrow{v}(B)_j$ for each $1\leq j<i$ and $\overrightarrow{v}(A)_i>\overrightarrow{v}(B)_i$. Thus, the leximin solution is a maximal element under $\succ$. We note that there could be multiple leximin solutions.

They observed that this solution is trivially PO, since if it were possible to improve the valuation of one agent without decreasing the valuation of any other agent, the new allocation would be strictly larger under $\succ$. 

\section{EFX: identical general valuations}\label{sec:efxident}

We start with the standard EFX property. Interestingly, this property might be incompatible with EFX$^{\pm}$ or PO even in problems where the agents' valuations are identical. This is either because EFX allocations might not exist or because they minimize the valuation of any agent.

\subsection{Problems with arbitrary items: impossibility}\label{subsec:mixedone}

We prove the \emph{first} major result. There are problems in our setting, where none of the allocations is EFX even under the non-zero marginal assumption. The key rationale behind this is the fact that these problems may contain mixed items. 

\begin{mytheorem}\label{thm:impefx}
There are problems with mixed items and identical general valuations whose marginals are non-zero, in which {\em no} allocation is EFX.
\end{mytheorem}  

\begin{myproof}
Consider a problem with \num{2} agents and \num{4} items. Define the valuations as follows: $v(\emptyset)=0$, $v(a)=5$, $v(b)=5$, $v(c)=5$, $v(d)=5$, $v(\lbrace a,b\rbrace)=6$, $v(\lbrace a,c\rbrace)=3$, $v(\lbrace b,c\rbrace)=3$, $v(\lbrace a,d\rbrace)=6$, $v(\lbrace b,d\rbrace)=6$, $v(\lbrace c,d\rbrace)=3$, $v(\lbrace a,b,c\rbrace)=7$, $v(\lbrace a,b,d\rbrace)=8$, $v(\lbrace a,c,d\rbrace)=7$, $v(\lbrace b,c,d\rbrace)=7$ and $v(\lbrace a,b,c,d\rbrace)=9$. Clearly, these valuations are identical. 

We claim that this is a problem with mixed items. To confirm this, we need to find a single mixed item. For this purpose, we consider the leximin solution $A=(\lbrace c\rbrace,\lbrace a,b,d\rbrace)$. In this allocation, item $a$ is bad for agent 1 wrt $A_1$ (i.e.\ $v(A_1\cup\lbrace a\rbrace)-v(A_1)=3-5=-2<0$) but good for agent 2 wrt $A_2$ (i.e.\ $v(A_2)-v(A_2\setminus\lbrace a\rbrace)=8-6=2>0$). This confirms our claim. 

We can now conclude that the problem is indeed with mixed items. We next show that none of the allocations in it is EFX. For this purpose, we consider all allocations in which agent 1 receive a different bundle and give one violation of this property per allocation. By the symmetry of the valuations, one can show that the corresponding allocations where agents swap bundles also violate EFX.

\begin{itemize}[noitemsep]
\item $A=(\emptyset,\lbrace a,b,c,d\rbrace)$: $v(A_1)=0$, $v(A_2)=9$, $v(A_2)-v(A_2\setminus\lbrace c\rbrace)=1$ and $v(A_1)=0<8=v(A_2\setminus\lbrace c\rbrace)$
\item $A=(\lbrace a\rbrace,\lbrace b,c,d\rbrace)$: $v(A_1)=5$, $v(A_2)=7$, $v(A_2)-v(A_2\setminus\lbrace c\rbrace)=1$ and $v(A_1)<6=v(A_2\setminus\lbrace c\rbrace)$
\item $A=(\lbrace b\rbrace,\lbrace a,c,d\rbrace)$: $v(A_1)=5$, $v(A_2)=7$, $v(A_2)-v(A_2\setminus\lbrace c\rbrace)=1$ and $v(A_1)<6=v(A_2\setminus\lbrace c\rbrace)$
\item $A=(\lbrace c\rbrace,\lbrace a,b,d\rbrace)$: $v(A_1)=5$, $v(A_2)=8$, $v(A_2)-v(A_2\setminus\lbrace d\rbrace)=2$ and $v(A_1)<6=v(A_2\setminus\lbrace d\rbrace)$
\item $A=(\lbrace d\rbrace,\lbrace a,b,c\rbrace)$: $v(A_1)=5$, $v(A_2)=7$, $v(A_2)-v(A_2\setminus\lbrace c\rbrace)=1$ and $v(A_1)<6=v(A_2\setminus\lbrace c\rbrace)$
\item $A=(\lbrace a,b\rbrace,\lbrace c,d\rbrace)$: $v(A_1)=6$, $v(A_2)=3$, $v(A_1)-v(A_1\setminus\lbrace a\rbrace)=1$ and $v(A_2)<5=v(A_1\setminus\lbrace a\rbrace)$
\item $A=(\lbrace a,c\rbrace,\lbrace b,d\rbrace)$: $v(A_1)=3$, $v(A_2)=6$, $v(A_2)-v(A_2\setminus\lbrace b\rbrace)=1$ and $v(A_1)<5=v(A_2\setminus\lbrace b\rbrace)$
\item $A=(\lbrace b,c\rbrace,\lbrace a,d\rbrace)$: $v(A_1)=3$, $v(A_2)=6$, $v(A_2)-v(A_2\setminus\lbrace d\rbrace)=1$ and $v(A_1)<5=v(A_2\setminus\lbrace d\rbrace)$
\myqed
\end{itemize}
\end{myproof}

This result compares favorably with the existence of EFX allocations in the case of identical additive valuations \cite{aleksandrov2020ki}. However, it remains an open problem whether such allocations exist with arbitrary (i.e.\ not just identical) additive utilities.

We also draw two interesting conclusions. First, the leximin solution might violate EFX in problems with mixed items and identical general valuations whose marginals are non-zero. Second, EFX and PO, or EFX and EFX$^{\pm}$ cannot generally be achieved in such problems.

\subsection{Problems without mixed items: incompatibility}\label{subsec:mixedthree}

One might hope that removing the mixed items in a given problem will help us restore some of the incompatibility results for EFX. However, our next pair of findings thwart this hope. They hold for problems with generally bad items and also for problems without mixed items (see Figure~\ref{fig:taxident}).

\begin{mytheorem}\label{thm:noefxstronger}
There are problems with generally bad items and identical general valuations whose marginals are non-zero, in which \emph{no} EFX allocation satisfies EFX$^{\pm}$.
\end{mytheorem}  

\begin{myproof}
Consider \num{2} agents, \num{4} items and the identical valuations: $v(\emptyset)=0$, $v(a)=-4$, $v(b)=-4$, $v(c)=-4$, $v(d)=-6$, $v(\lbrace a,b\rbrace)=-5$, $v(\lbrace a,c\rbrace)=-5$, $v(\lbrace b,c\rbrace)=-5$, $v(\lbrace a,d\rbrace)=-7$, $v(\lbrace b,d\rbrace)=-7$, $v(\lbrace c,d\rbrace)=-7$, $v(\lbrace a,b,c\rbrace)=-8$, $v(\lbrace a,b,d\rbrace)=-8$, $v(\lbrace a,c,d\rbrace)=-8$, $v(\lbrace b,c,d\rbrace)=-8$ and $v(\lbrace a,b,c,d\rbrace)=-9$. 

We argue that the way to achieve the axiomatic quarantees of EFX is to give $\lbrace a,b,c\rbrace$ to agent 1 and $\lbrace d\rbrace$ to agent 2, or to swap these bundles. For each other allocation, one can find a violation of this property as in Theorem~\ref{thm:impefx}. However, the allocation $A=(\lbrace a,b,c\rbrace,\lbrace d\rbrace)$ violates EFX$^{\pm}$: $v(A_1)=-8<-7=v(A_2\cup\lbrace a\rbrace)$.
\myqed
\end{myproof}

We conclude that the set of EFX allocations and the set of EFX$^{\pm}$ allocations might be disjoint in some problems even under the non-zero marginal assumption. As a result, there are problems where none of the EFX$^{\pm}$ allocations is EFX. 

Further, Plaut and Roughgarden \shortcite{plaut2018} proved that the leximin solution is EFX and PO in problems with general but identical valuations for goods (i.e.\ generally good items), subject to the non-zero marginal assumption. Interestingly, adding bads to the problem may destroy this compatibility. 

\begin{mytheorem}\label{thm:noefxpostronger}
There are problems with generally bad items and identical general valuations whose marginals are non-zero, in which \emph{no} EFX allocation satisfies PO.
\end{mytheorem}  

\begin{myproof}
Consider again the problem from Theorem~\ref{thm:noefxstronger}. The only two EFX allocations in this problem are $A=(\lbrace d\rbrace,\lbrace a,b,c\rbrace)$ and $B=(\lbrace a,b,c\rbrace,\lbrace d\rbrace)$. Pick also the allocations $C=(\lbrace a,b\rbrace,\lbrace c,d\rbrace)$ and $D=(\lbrace c,d\rbrace,\lbrace a,b\rbrace)$. 

Each of these is the leximin solution. Further, we have that $v(C_1)=-5>-6=v(A_1)$, $v(C_2)=-7>-8=v(A_2)$, $v(D_1)=-7>-8=v(B_1)$, $v(D_2)=-5>-6=v(B_2)$ hold. Hence, $C$/$D$ Pareto-improves $A$/$B$.
\myqed
\end{myproof}

This result has the following insightful implication. That is, there are problems where all EFX allocations minimize the maximum valuation of any agent. This suggests that any approach that maximizes the minimum valuation of any agent might fail to deliver any EFX guarantees. 

\section{EFX$^{\pm}$ and PO: identical general valuations}\label{sec:refxident}

We further analyse the new EFX$^{\pm}$ property. We prove the {\em second} major result. That is, EFX$^{\pm}$ and PO allocations exist in each problem in our setting as long as the agents' valuations are general but identical. For example, the leximin solution is EFX$^{\pm}$ and PO. 

\begin{mytheorem}\label{thm:newefxpo}
With general but identical valuations, the leximin solution satisfies EFX$^{\pm}$ and PO.
\end{mytheorem}  

\begin{myproof}
Let $A$ be an leximin$++$ allocation. This allocation is PO even with general (and not necessarily identical) valuations. For this reason, we next show that $A$ is EFX$^{\pm}$. Assume that $A$ is not EFX$^{\pm}$ for a pair of agents $i,j\in [n]$ with $i\neq j$. That is, $v(A_i)<v(A_j)$. 

For our proof, we let $v(A_1)\leq\ldots\leq v(A_n)$ denote the utility order induced by $A$. We also let $k=arg\max \lbrace h\in [n]|v(A_h)\leq v(A_i)\rbrace$. We note that $i\leq k$ and $k<j$ hold. Thus, we can conclude that $v(A_i)=v(A_k)$ and $v(A_k)<v(A_j)$ hold. 

The violation of EFX$^{\pm}$ further means that at least one of the following two conditions should hold: (a) $\exists o\in A_j:v(A_i\cup\lbrace o\rbrace)>v(A_i),v(A_i)<v(A_j\setminus \lbrace o\rbrace)$ and (b) $\exists o\in A_i:v(A_i)<v(A_i\setminus\lbrace o\rbrace),v(A_i)<v(A_j\cup\lbrace o\rbrace)$. We consider two cases depending on which of (a) or (b) holds.

\emph{Case 1}: If (a) holds for some $o\in A_j$, then let us move only item $o$ from $A_j$ to $A_i$. We let $B$ denote this allocation: $B_i=A_i\cup\lbrace o\rbrace$, $B_j=A_j\setminus\lbrace o\rbrace$ and $B_h=A_h$ for each $h\in [n]\setminus\lbrace i,j\rbrace$. We next prove that $B\succ A$ holds. 

Wlog, let $v(B_{p_1})\leq\ldots\leq v(B_{p_n})$ denote the utiity order induced by $B$. We note that the positions of agents $i$ and $j$ in this order are at least $k$. As a result, $B_{p_q}=A_q$ for each $q\in\lbrace 1,\ldots, k\rbrace\setminus\lbrace i\rbrace$. 

We now consider three cases for the $k$th agent in this order. If $p_k=i$, we derive $v(B_i)=v(A_i\cup\lbrace o\rbrace)>v(A_i)$ by (a). If $p_k=j$, we also derive $v(B_j)=v(A_j\setminus\lbrace o\rbrace)>v(A_i)$ by (a). If $p_k=k+1$, we conclude $v(B_{k+1})=v(A_{k+1})>v(A_k)\geq v(A_i)$ by the construction of $B$ and the choice of $k$. 

To conclude this case, we simple observe that $v(B_{p_q})\geq v(B_{p_k})>v(A_i)$ holds for each $q\in\lbrace k+1,\ldots,n\rbrace$. Gathering the pieces together, it follows that $A$ cannot be the leximin solution. This is a contradiction. Hence, (a) cannot hold. 

\emph{Case 2}: If (b) holds for some $o\in A_i$, then let us move only item $o$ from $A_i$ to $A_j$. We let $B$ denote this allocation: $B_i=A_i\setminus\lbrace o\rbrace$, $B_j=A_j\cup\lbrace o\rbrace$ and $B_h=A_h$ for each $h\in [n]\setminus\lbrace i,j\rbrace$. We again prove that $B\succ A$ holds. 

Consider again the utiity order induced by $B$, say $v(B_{p_1})\leq\ldots\leq v(B_{p_n})$. The positions of agents $i$ and $j$ in this order are also at least $k$. As a result, $B_{p_q}=A_q$ for each $q\in\lbrace 1,\ldots, k\rbrace\setminus\lbrace i\rbrace$. 

The cases for the $k$th agent are similar as in the first case. If $p_k=i$, we derive $v(B_i)=v(A_i\setminus\lbrace o\rbrace)>v(A_i)$ by (b). If $p_k=j$, we also derive $v(B_j)=v(A_j\cup\lbrace o\rbrace)>v(A_i)$ by (b). If $p_k=k+1$, we conclude $v(B_{k+1})=v(A_{k+1})>v(A_k)\geq v(A_i)$ by the construction of $B$ and the choice of $k$. 

At the end, we again observe that $v(B_{p_q})\geq v(B_{p_k})>v(A_i)$ holds for each $q\in\lbrace k+1,\ldots,n\rbrace$. Consequently, the allocation $A$ cannot be the leximin solution. This leads again to a contradiction. Therefore, (b) also cannot hold.
\myqed
\end{myproof}

\section{EFX$^{\pm}$: general valuations}\label{sec:refxefxgen}

It is well-known that the egalitarian allocation, maximizing the minimum valuation of any agent, might fail EF1 in problems with \num{3} agents and additive valuations \cite{caragiannis2016}. This holds for the leximin solution and EF1$^{\pm}$ in this case. For this reason, we study the case of \num{2} agents.

\subsection{The case of $2$ agents}\label{subsec:two agents}

We come to the {\em third} major result. This one is for general but \emph{disjointly normalised} valuations. That is, for each $M,N\subseteq [m]$ such that $M\cap N=\emptyset$ and $M\cup N=[m]$, we have $v_i(M)+v_i(N)=c$ for each $i\in [2]$ and some $c\in\mathbb{R}$. Interestingly, each problem with such valuations admits an EFX$^{\pm}$ and PO allocation. 

\begin{mytheorem}\label{thm:newefxpotwo}
With general (not just identical) but disjointly normalised valuations, the leximin solution satisfies EFX$^{\pm}$ and PO allocation.
\end{mytheorem}  

\begin{myproof}
Let $A$ be the leximin solution. Clearly, $A$ is PO. Suppose that $A$ is not EFX$^{\pm}$. Wlog, let agent 1 be not EFX$^{\pm}$ of agent 2. Hence, it must be the case that (a) $v_1(A_1)<v_1(A_2\cup\lbrace o\rbrace)$ holds for some $o\in A_1$ with $v_1(A_1\setminus\lbrace o\rbrace)>v_1(A_1)$ or (b) $v_1(A_1)<v_1(A_2\setminus\lbrace o\rbrace)$ holds for some $o\in A_2$ with $v_1(A_1\cup\lbrace o\rbrace)>v_1(A_1)$. We consider two cases.

If (a) holds for $o\in A_1$, let us consider bundles $S_1=A_1\setminus\lbrace o\rbrace$ and $S_2=A_2\cup\lbrace o\rbrace$. If (b) holds for $o\in A_2$, let us consider bundles $S_1=A_1\cup\lbrace o\rbrace$ and $S_2=A_2\setminus\lbrace o\rbrace$. We construct an allocation $B$ that leximin-dominates $A$, reaching a contradiction. We let $B_1=\argmin_S v_2(S)$ and $B_2=\argmax_S v_2(S)$ where $S\in\lbrace S_1,S_2\rbrace$.

By construction, $v_2(B_2)\geq v_2(B_1)$ holds in $B$. Moreover, $v_1(S_1)>v_1(A_1)$ and $v_1(S_2)>v_1(A_1)$ hold in each of the cases (a) and (b). These inequalities follow because agent 1's marginal valuations for the moved item are non-zero and agent 1 is not EFX$^{\pm}$ of agent 2. We conclude $v_1(B_1)>v_1(A_1)$.

We have $v_1(A_1)+v_1(A_2)=c$ and $v_2(A_1)+v_2(A_2)=c$ for some $c\in \mathbb{R}$ by $A_1\cap A_2=\emptyset$ and the fact that the valuations are disjointly normalised. As $v_1(A_1)<v_1(A_2)$, it follows $v_1(A_1)<c/2$. By the PO of $A$, $v_2(A_2)>v_2(A_1)$. Hence, $v_2(A_1)<c/2$ and $v_2(A_2)>c/2$. This implies $v_1(A_1)<v_2(A_2)$.

Further, the allocation $B$ is also such that $B_1\cap B_2=\emptyset$ holds. Therefore, $v_2(B_1)+v_2(B_2)=c$. As $v_2(B_2)\geq v_2(B_1)$, it follows that the inequality $v_2(B_2)\geq c/2$ holds. We now derive the contradiction: $\min\lbrace v_1(A_1),v_2(A_2)\rbrace=v_1(A_1)<\min\lbrace v_1(B_1),c/2\rbrace$ $\leq\min\lbrace v_1(B_1),v_2(B_2)\rbrace$.
\myqed
\end{myproof} 

A special but very common and practical sub-case of disjointly normalised valuations is the one when the valuations are additive and {\em normalised}, i.e.\ $v_i([m])=c$ for $i\in\lbrace 1,2\rbrace$ and some $c\in\mathbb{R}$. Nevertheless, we might wish to drop the assumption of disjointly normalised valuations and achieve just EFX$^{\pm}$, or even additionally PO. 

On the plus side, we can give a ``cut-and-choose'' protocol for returning an EFX$^{\pm}$ allocation: (1) agent 1 ``cut'' the bundle of items in two, using the leximin solution and supposing that agent 2 have identical valuations, and (2) agent 2 ``choose'' their most favorable bundle. A similar idea was used by Plaut and Roughgarden \shortcite{plaut2018} for EFX and goods.

By Theorem~\ref{thm:newefxpo}, it follows that agent 1 is EFX$^{\pm}$ for each bundle after the cut. As agent 2 pick their most favorable bundle, they are envy-free of agent 1. This concludes our claim. The interesting part about this simple result in contrast to the one in Theorem~\ref{thm:newefxpotwo} is that agents can have general valuations that might not necessarily be disjointly normalised.

\begin{mycorollary}\label{cor:one}
With \num{2} agents and general (not just identical) valuations, an EFX$^{\pm}$ allocation exists.
\end{mycorollary}

On the minus side, both properties might be incompatible. Plaut and Roughgarden \shortcite{plaut2018} proved a similar result for EFX$_0$ and PO in problems with \num{2} agents and generally good items under the non-zero marginal assumption. We simply note that EFX$_0$ coincides with our EFX$^{\pm}$ in this case because of the non-zero marginal assumption.

\begin{mycorollary}\label{cor:two}
There are problems with generally good items and general but not identical valuations whose marginals are non-zero, in which \emph{no} EFX$^{\pm}$ allocation satisfies PO.
\end{mycorollary}

\section{EF1 and PO: general valuations}\label{sec:efonepogen}

We turn attention to the weaker EF1 property. We note that the compatibility between EF1 and PO remains an open question with additive valuations. However, there are very simple problems in our setting where each EF1 allocation violates PO. This is our \emph{fourth} major result.

\begin{mytheorem}\label{thm:noefonepo}
There are problems with arbitrary items and general but not identical valuations whose marginals are non-zero, in which \emph{no} EF1 allocation satisfies PO.
\end{mytheorem}  

\begin{myproof}
Consider \num{2} agents and \num{4} items. We define the valuations as follows: $v_1(\emptyset)=v_2(\emptyset)=0$, $v_1(S)=v_2(S)=3$ for $S$ with $|S|=3$ and $v_1(S)=v_2(S)=4$ for $S$ with $|S|=4$. For $S$ with $|S|=1$, $v_1(S)=-1$ and $v_2(S)=1$. For $S$ with $|S|=2$, $v_1(S)=-2$ and $v_2(S)=2$. 

To achieve EF1, agent 1 cannot get a bundle with \num{3} or \num{4} items. If they got such a bundle, then agent 2 got a bundle with at most \num{1} item. But, then agent 2's valuation would be at most \num{1} and agent 2's valuation for agent 1's bundle be at least \num{3}. Removing an item from agent 1's bundle or agent 2's bundle would not eliminate the envy of agent 2.

Consequently, it must be the case that agent 1 get zero, one or two items. If they got no item, their valuation for their own bundle would be \num{0} and their valuation for the bundle of agent 2 be \num{4}. Again, removing an item from agent 2's bundle would not eliminate this envy because $v_1(S)=3$ for each $S$ with $|S|=3$.

Hence, agent 1 should get one item or two items. If they got one item, the agents' valuations are $-1$ and $3$. If they got two items, the agents' valuations are $-2$ and $2$. We note that each of these is EF1. However, it is easy to see that these are Pareto dominated by the valuations $0$ and $4$, i.e.\ agent 1 get no item and agent 2 get all items.   
\myqed
\end{myproof}

This result gives us another reason to focus on EF1$^{\pm}$. The other reason for this was that it is unknown whether EF1 allocations exist in problems with \num{3} or more agents and identical general valuations \cite{brczi2020envyfree}. By comparison, EF1$^{\pm}$ and PO allocations exist in this case by Theorem~\ref{thm:newefxpo}.

\section{EF1$^{\pm}$: general valuations}\label{sec:newefonegen}

We lastly sum up some observations for the new EF1$^{\pm}$ property. We observe that each EF1 allocation in the problem from Theorem~\ref{thm:noefonepo} also satisfies EF1$^{\pm}$ and no other allocation is EF1$^{\pm}$. It follows immediately that EF1$^{\pm}$ and PO might be incompatible in some problems in our setting. 

\begin{mycorollary}\label{cor:three}
There are problems with arbitrary items and general but not identical valuations whose marginals are non-zero, in which \emph{no} EF1$^{\pm}$ allocation satisfies PO.
\end{mycorollary}

We note that this impossibility result is in-line with the possibility result in Theorem~\ref{thm:newefxpotwo} simply because the valuations in the problem from Theorem~\ref{thm:noefonepo} are not disjointly normalised. Therefore, the two results do not contradict but complement each other.

By Corollary~\ref{cor:one}, an EF1$^{\pm}$ allocation also exists in each problem with \num{2} agents and general valuations. We believe that this complements nicely the recent possibility result of Bérczi et al.\ \shortcite{brczi2020envyfree} who proved that an EF1 allocation exists in each such problem. 

Finally, Aziz et al. \shortcite{aziz2019gc} presented the double round-robin algorithm for returning an EF1 allocation in problems with additive valuations. In such problems, recal that an allocation is EF1 iff it is EF1$^{\pm}$. Hence, their algorithm returns an EF1$^{\pm}$ allocation in this case. 

\section{Discussion}\label{sec:disc}

We considered a stronger variant of EFX$^{\pm}$ where the agents' marginal valuations are allowed to be zero, say EFX$^{\pm}_0$. With additive valuations for goods, this is EFX$_0$ from \cite{kyropoulou2019}. With additive valuations for bads, it is EFX$^+_0$ from \cite{brczi2020envyfree}. Aleksandrov and Walsh \shortcite{aleksandrov2020ki} generalized EFX$_0$ to the case of additive valuations for goods and bads when it coincides with EFX$^{\pm}_0$. Thus, they gave problems with \num{2} agents and identical additive valuations where none of the allocations satisfies EFX$^{\pm}_0$. 

We also considered two other variants of envy-freeness up to any item. The first one uses condition 1)$^{\prime}$ in the definition of EFX$^{\pm}$ and condition 2) in the definition of EFX. The impossibility result in Theorem~\ref{thm:impefx} holds for this variant. The second variant uses condition 1) in the definition of EFX and condition 2)$^{\prime}$ in the definition of EFX$^{\pm}$. The incompatibility result in Theorem~\ref{thm:noefxpostronger} holds for this variant. This perhaps suggests that EFX$^{\pm}$ is the ``right'' notion for our model.

Finally, we observed that the variant of EFX from \cite{chen2020fairness} cannot be combined with PO even in problems with generally good items. To see this, consider \num{2} agents and the identical valuations $v(\lbrace a,b\rbrace)=2$, $v(a)=1$, $v(b)=0$ and $v(\emptyset)=0$. To achieve PO, we should give $\lbrace a,b\rbrace$ to one of the agents. Wlog, let $A=(\lbrace a,b\rbrace,\emptyset)$. This allocation violates Chen and Liu's variant of EFX: $v(A_1)-v(A_1\setminus\lbrace b\rbrace)=1>0$ and $v(A_2)=0<1=v(A_1\setminus\lbrace b\rbrace)$.

\section{Conclusion}\label{sec:conc}

We considered a fair division model in which agents have general valuations for bundles of items. We proposed two new axiomatic properties for allocations in this model: EF1$^{\pm}$ and EFX$^{\pm}$. We compared these with two existing properties: EF1 and EFX. Table~\ref{tab:results} contains our results. Although EF1 and EF1$^{\pm}$ allocations exist with \num{2} agents, these results suggest that EFX$^{\pm}$ and PO allocations are compatible in each case where EFX and PO allocations are incompatible. We also proved some interesting impossibility results: (1) an EFX allocation might not exist even under the non-zero marginal assumption and (2) an EF1 and PO allocation might also not exist in general.

\bibliographystyle{named}
\bibliography{efx}

\end{document}